\documentclass[]{spie}  

\usepackage{amsmath,amsfonts,amssymb}
\usepackage{caption}
\usepackage{subcaption}
\usepackage{graphicx}
\usepackage[colorlinks=true, allcolors=blue]{hyperref}
\usepackage{siunitx}

\title{Modeling sidelobe response for ground-based mm-wavelength telescopes with the geometrical theory of diffraction}
\author[1,*]{Alexandre E. Adler}
\author[1]{Jon E. Gudmundsson}

\affil[1]{{The Oskar Klein Centre, Department of Physics, Stockholm University, SE-106 91 Stockholm, Sweden}}

\authorinfo{Further author information: Send correspondence to Alexandre E. Adler, E-mail: alexandre.adler@fysik.su.se}

\newcommand{\mrm}[1]{\mathrm{#1}} 

\usepackage{xcolor}

\newcommand\lcdm{$\Lambda$CDM }

\DeclareSIUnit{\beli}{Bi}
\DeclareSIUnit{\dBi}{\deci\beli}

\begin{document} 
\maketitle

\begin{abstract}
Accurate optical modeling is important for the design and characterisation of current and next-generation experiments studying the Cosmic Microwave Background (CMB).
Geometrical Optics (GO) cannot model diffractive effects. 
In this work, we discuss two methods that incorporate diffraction, Physical Optics (PO) and the Geometrical Theory of Diffraction (GTD). 
We simulate the optical response of a ground-based two-lens refractor design shielded by a ground screen with time-reversed simulations. 
In particular, we use GTD to determine the interplay between the design of the refractor's forebaffle and the sidelobes caused by interaction with the ground screen.

\end{abstract}


\section{INTRODUCTION}
\label{sec:intro}  
The temperature and polarization anisotropies of the Cosmic Microwave Background (CMB) have been studied for now nearly thirty years with ever-increasing precision by space \cite{Smoot92, Bennett2013, Planck2018_overview}, balloon \cite{Crill2003, Hanany2000, Fraisse2013}, and ground-based\cite{ACTPol2016, BICEP2} millimetre telescopes. 
Temperature anisotropies are variations of a few tens of $\mu K_\mathrm{CMB}$, polarization anisotropies are an order of magnitude weaker. 
Increasingly precise knowledge of the CMB's anisotropies has produced some of the most stringent constraints on our \lcdm cosmological model.\cite{} 
This has been experimentally enabled by concurrent advances in cryogenics, electronics, detector design and optics. 
As instruments become more sensitive, the impact of various systematic effects grows. Improving the accuracy of our optical models represents one key aspect of the overall effort to control systematic effects.

Optical systems can be modeled using a wide range of techniques. A common and relatively computationally inexpensive approach is known as geometrical optics (GO), or more commonly as ray tracing. This method can be used to quickly identify key properties of common optical systems. However, geometrical optics does not account for diffraction effects, and therefore this approach is unable to capture some features in the far-field response of these telescopes.

Current and upcoming experiments searching for faint hypothesized primordial B-modes in the polarization of the CMB require exquisite control of instrument non-idealities. These CMB polarimeters typically employ some kind of optical baffling to shield the primary optics from unwanted sources. Unfortunately, diffraction caused by such baffling elements can redistribute optical power to wide angles, thereby solving one problem but introducing another. In order to properly quantify such trade-offs, we need optical simulations that account for the wavelike properties of light; this includes a commonly used method known as physical optics (PO). Some baffling elements however, such as satellite V-grooves or the screens used by ground-based telescopes, tend to be electrically large, thereby hindering the use of physical optics simulations in a reasonable computation time. 

In this work, we combine physical optics simulations with a modified ray-tracing method known as the Geometrical Theory of Diffraction (GTD) \cite{Keller1956, Keller1959, Keller1962} to predict the far field beam response of a fiducial ground-based telescope. Using a Python API that communicates with a commercial software named TICRA Tools (formerly GRASP) \cite{TicraTools}, we generate an ensemble of PO and GTD simulations.\footnote{See \url{https://www.ticra.com/}} Similar approaches are described in Refs.~\citenum{Sandri2004},~\citenum{Tauber2010},~and~\citenum{Qubic20-8}.

This proceeding is organised as follows: Section \ref{sec:math} introduces the mathematical and physical framework for PO and GTD. 
Section~\ref{sec:instrument} describes our instrument design and the parameters for our baffling elements for GTD simulations of diffraction by a ground screen. 
The simulation results are then presented and discussed in Section~\ref{sec:results}. 
We conclude in Section~\ref{sec:conclusion}.

\section{BASICS OF PHYSICAL OPTICS AND GTD}
\label{sec:math}
\subsection{Physical optics}
Physical Optics is a method that allows one to sequentially propagate electric fields scattered on surfaces with arbitrary geometries and material properties. An incoming electric field on that surface will induce surface currents, which will themselves induce a radiated electric field. 
The scattered electric field is then the sum of the incoming field and the radiated one. 
Physical optics approximates the current induced by the incoming field at a point on a curved surface with the current induced at the same point on an infinite flat plane tangent to the curved surface: 
\begin{equation}
    {\bf J}_\textrm{e} = 2 {\bf\hat n}\times {\bf H}_\mathrm{inc},
\end{equation}
where ${\bf J}_\textrm{e}$ is the induced current, ${\bf\hat n}$ is the surface normal, and ${\bf H}_\textrm{inc}$ is the magnetic field of the incoming radiation. For regions that are not illuminated by the source, the induced currents are assumed to be zero. 
By suitably sampling the surface, it is thus possible to establish the field scattered by the object at arbitrary locations.
The spacing of the integration points needs to be smaller than the wavelength of the illuminating radiation. 
In some cases, assumptions of azimuthal symmetry can reduce the number of integration points, however this simplification is not valid for many systems observing at mm-wavelengths. When objects become electrically large, PO simulations become computationally intensive: in those cases, we require alternative methods to make modeling tractable.

\subsection{The geometrical theory of diffraction: expanding from the Sommerfeld solution}

In geometrical optics, rays travel along straight lines until they either reflect or refract. Snell's law and the Fresnel equations describe the rays that result from these interactions. In addition to the laws of GO, GTD introduces rules to create diffracted fields in shadow regions: for instance in the shadow formed when an aperture truncates a bundle of rays. 
GTD creates extra rays to bridge over the discontinuity in that boundary region. 
Furthermore, these diffracted rays are generated only by rays incident on discontinuities of the object, the so-called edge rays that delineate the boundary between light and shadow in GO. 
The amplitude of the field for the diffracted rays is proportional to the amplitude of the diffracted ray at the edge multiplied by a diffraction coefficient $D(\mathbf{n}_1, \mathbf{n}_2)$ where $\mathbf n_1$ and $\mathbf n_2$ are the unit vectors of the incident and diffracted rays.
The best known example of GTD is the Sommerfeld solution for diffraction on a reflecting half-plane.

\begin{figure}
    \centering
    \includegraphics[width=\textwidth]{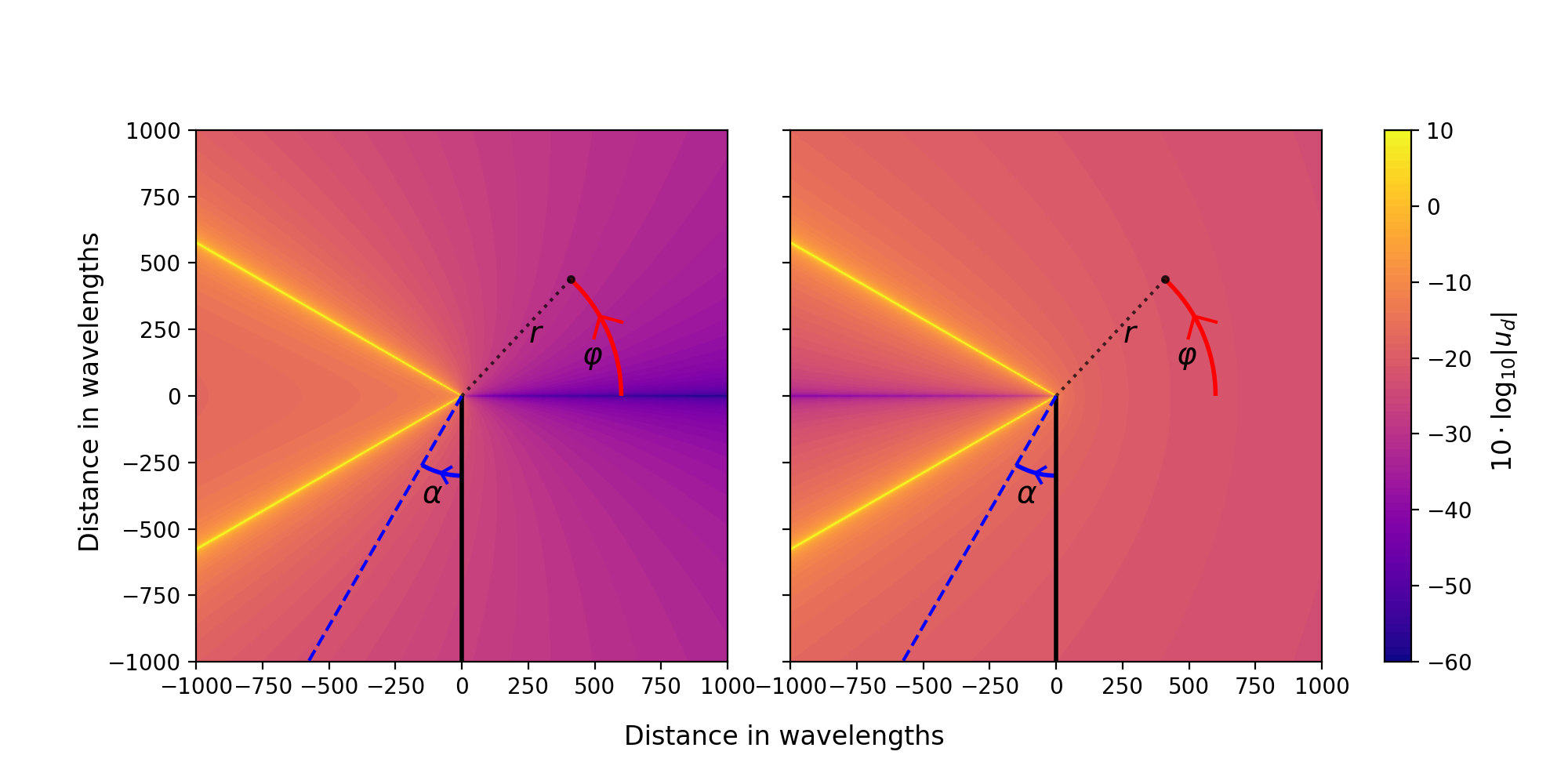}
    \caption{Modulus of the relative diffracted amplitude following Equation~\ref{eq:Sommerfeld} for a range of $(r,\varphi)$. The incoming plane wave is assumed to have incidence $\alpha=\pi/6$. The angle conventions for $\alpha$ and $\varphi$ are labeled in blue and red respectively. Left (right) panel is for the electric field parallel (perpendicular) to the plane of incidence. Note the caustics in each panel, as the solution diverges when $\varphi \pm \alpha = \pi$.}
    \label{fig:Sommerfeld}
\end{figure}

We define a reflective half-plane along the $\varphi=-\pi/2$ line in polar coordinates $(r,\varphi)$, and propagate a plane wave with wavenumber $k$ so that the wave front is incident on the half-plane reflector with angle $\alpha$. The angle conventions are shown in the Fig.~\ref{fig:Sommerfeld}. The diffracted field amplitude at a distance $r>\!\!>k^{-1}$ from the edge is then:
\begin{equation}
   u_{\alpha}(r,\varphi) = u_0 e^{i(kr+\frac{\pi}{4})}\frac{1}{2\sqrt{2\pi kr}}f_s(\alpha, \varphi) = \frac{u_0 e^{i(kr+\frac{\pi}{4})}}{2\sqrt{2\pi r}}D_s(k, \alpha, \varphi)\textrm{,} \label{eq:Sommerfeld}
\end{equation}
where $u_0$ the amplitude of the field at the edge of the screen, $D_s=k^{-1/2}f_s$ is the diffraction coefficient, and $f_s$ contains the angular dependence:
\begin{equation}
    f_s(\alpha, \varphi) =  -\frac{1}{\cos{\frac{\varphi - \alpha}{2}}} \mp \frac{1}{\cos{\frac{\varphi+\alpha}{2}}}  \: \textrm{.} \label{eq:geodiff}
\end{equation}
The second term in the above equation is negative (positive) if the electric field is parallel (perpendicular) to the edge of the semi-infinite plane. The solution is not applicable when $\varphi = \pi \pm \alpha$, as this creates unphysical caustics in the field. This expression, derived by Sommerfeld \cite{Sommerfeld96}, accurately describes diffraction on sharp edges. Similar expressions have been worked out by Keller et al. for diffraction on rounded edges or wedges\cite{Keller1956, Keller1959, Keller1962}. 

The module that provides GTD simulation capabilities in TICRA Tools is called GRASP. Unfortunately, GRASP provides a version of the GTD approach that only includes Sommerfeld's solution. We are therefore currently unable to use its GTD implementation to model diffraction on rounded edges. As expected, however, the semi-analytical expressions derived for diffraction on rounded edges generally predicts diffraction patterns that have lower amplitude compared to ones found for diffraction on a sharp edge by Sommerfeld. Since GRASP only implements Sommerfeld's solution, we expect that our results represent upper limits compared to experimental configurations that implement rounded edges. We will end this section by describing how the GRASP GTD procedure works, to explicit its computational edge over PO.

The GTD method implemented by GRASP starts the set of points in space on which the output will be computed. 
From these points, rays are propagated towards the light source and any specified surface. 
Rays that encounter a surface get reflected, either towards the source directly or towards another reflector in the GTD system. 
Rays that encounter an edge will generate extra rays following the rules of GTD. 
The total number of reflections and diffraction steps allowed for a given input ray is an adjustable parameter that represents the maximum order of the simulation.
Once the paths from source to output are established, the field can be computed by adding up the rays, much like in GO.
The obvious advantage over PO comes from not having to compute the equivalent currents in the objects on which reflections and diffractions occur, using the computationally easier formulations of GTD. This is beneficial for electrically large objects such as the surface of a ground screen. The GTD implementation in GRASP allows surfaces to block ray paths, but ignores partial refraction. 
In that respect, PO analysis is superior.

\section{INSTRUMENT SETUP}
\label{sec:instrument}
\subsection{Telescope design}
\begin{figure}
    \centering
    \begin{subfigure}[b]{0.3\textwidth}
        \includegraphics[width=\textwidth]{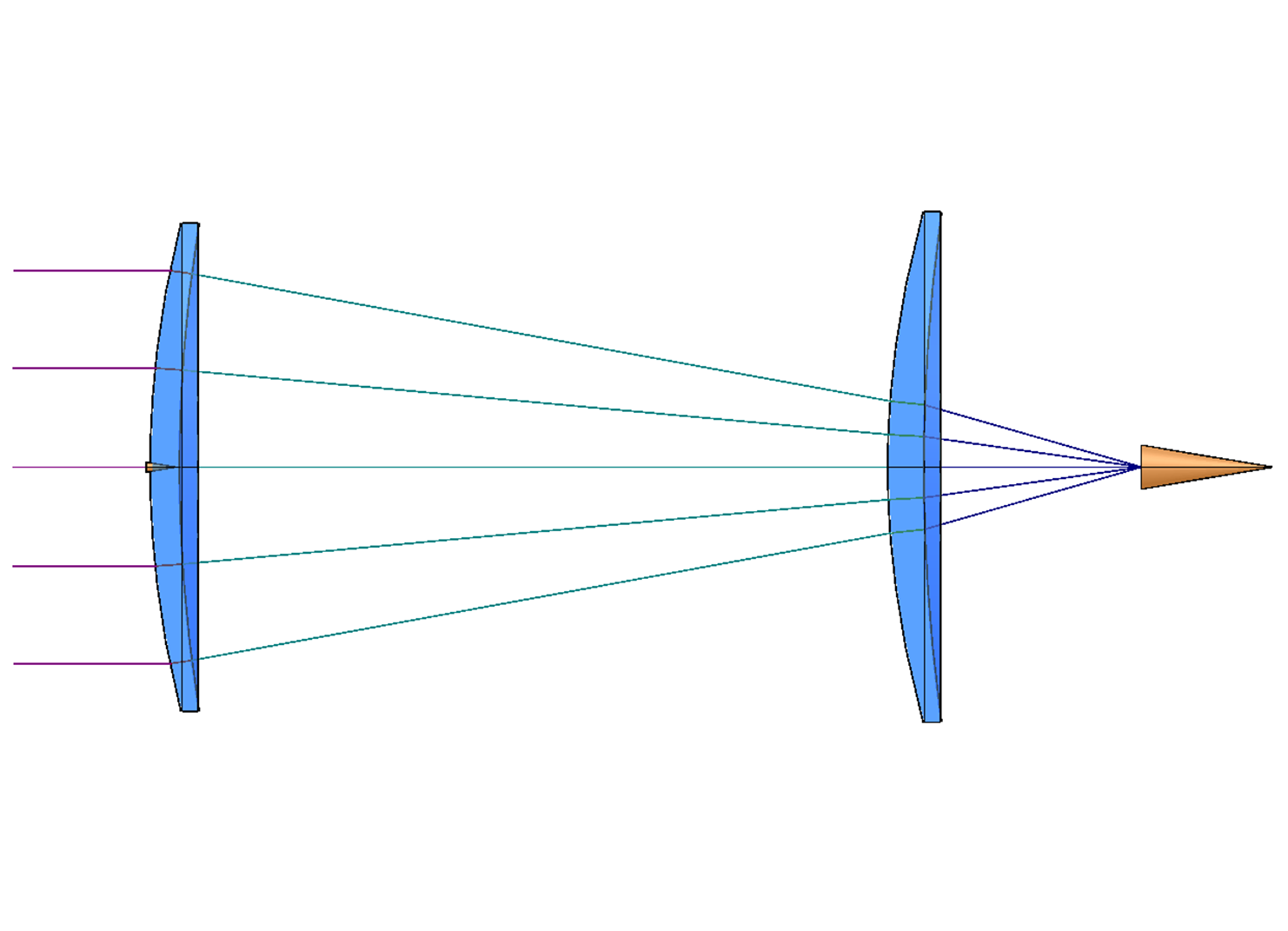}
         \caption{Lenses and Gaussian feedhorn, rays traced to \ang{16} from axis.}
         \label{fig:lens}
    \end{subfigure}
    \hfill
    \begin{subfigure}[b]{0.3\textwidth}
        \includegraphics[width=\textwidth]{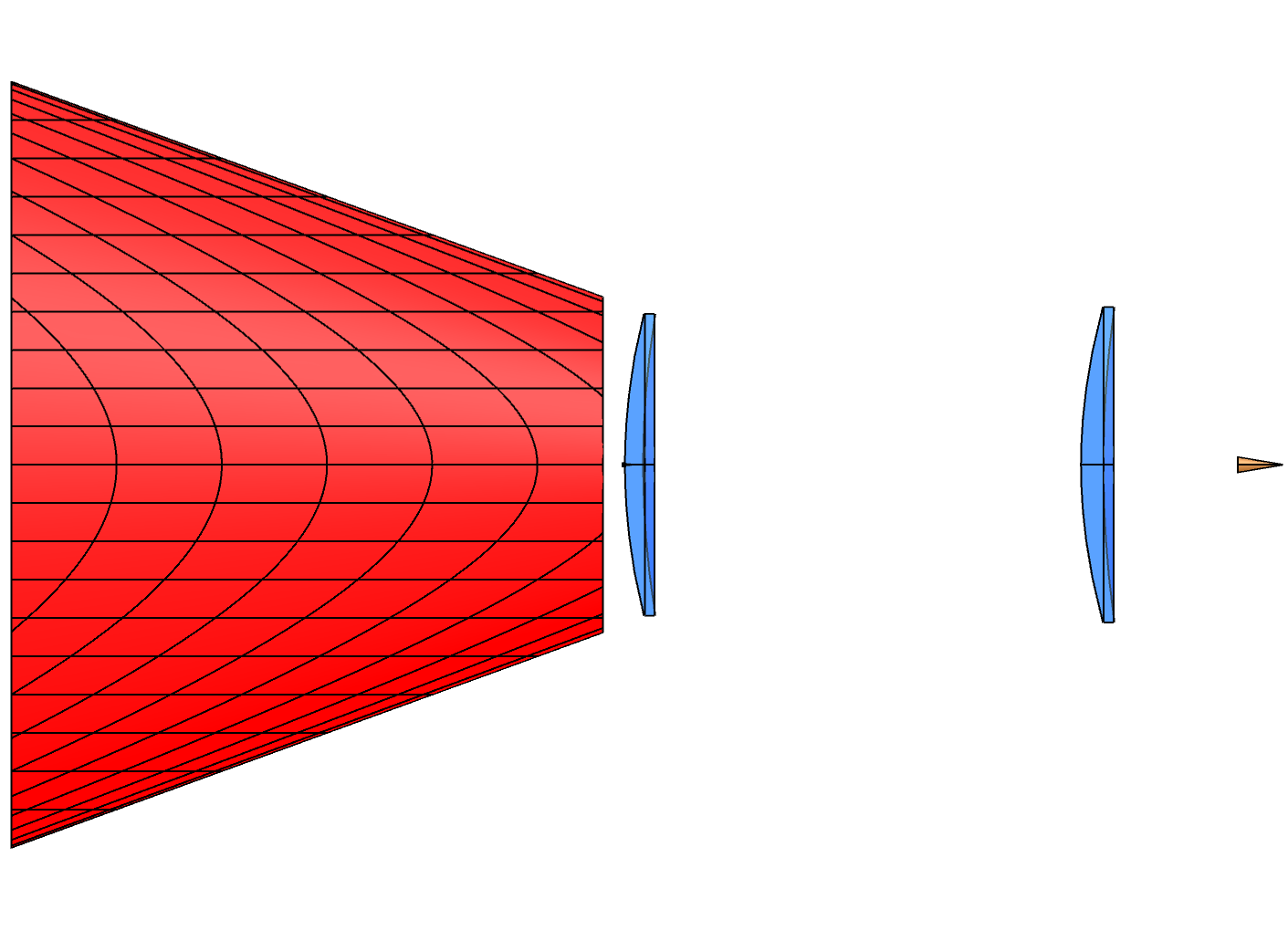}
         \caption{Lenses, and Gaussian feedhorn, Lambertian source and forebaffle}
         \label{fig:lfb}
    \end{subfigure}
    \hfill
    \begin{subfigure}[b]{0.3\textwidth}
        \includegraphics[width=\textwidth]{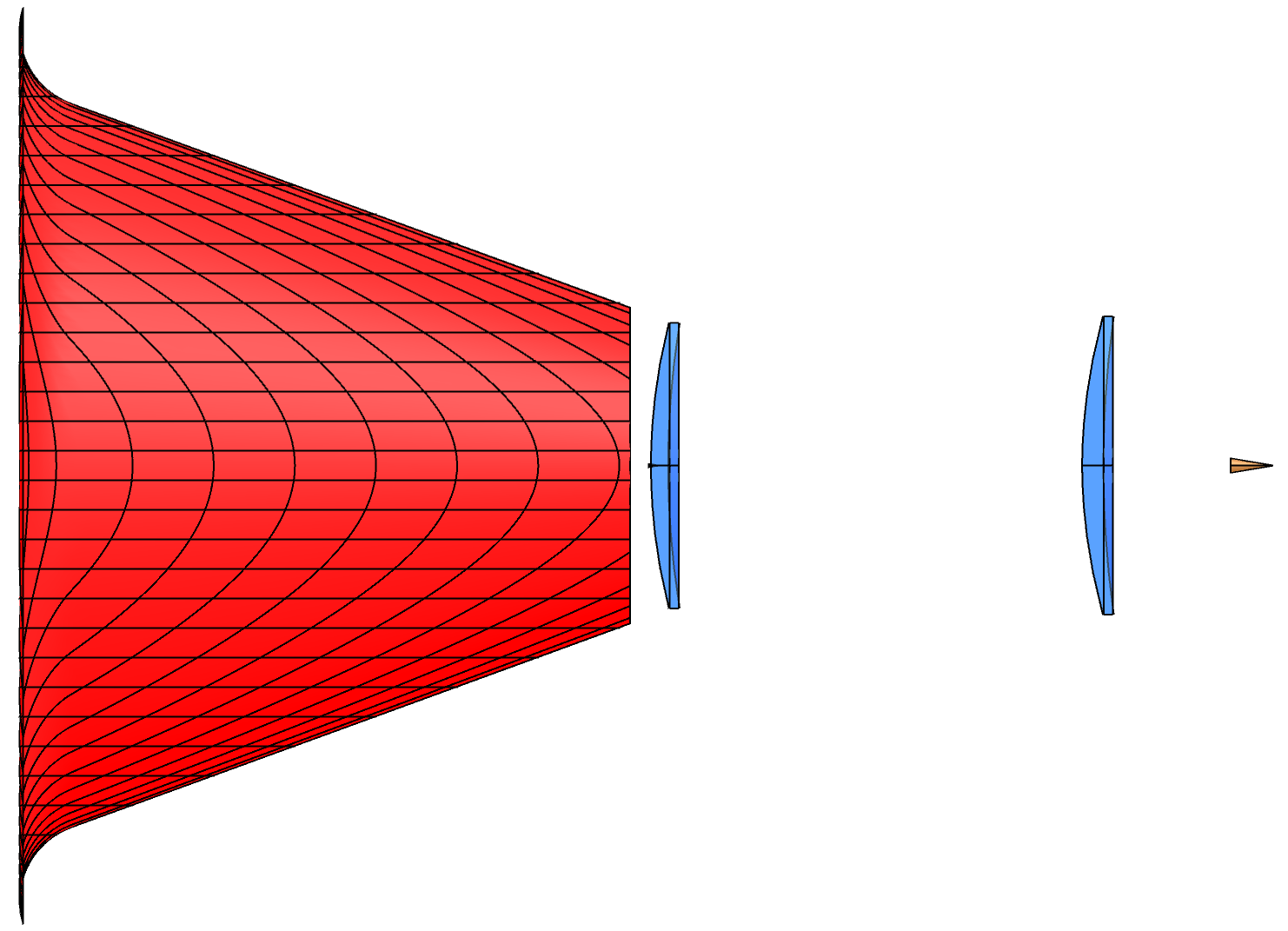}
         \caption{Lenses, source horn, Lambertian source, forebaffle with a curved rim}
         \label{fig:lfbc}
    \end{subfigure}
    \caption{Parts of the system used for the PO analysis}
    \label{fig:forebaffle}
\end{figure}

We study a simple two-lens silicon refractor telescope, first described in Ref~\citenum{Gudmundsson2020}, that could be used to observe the degree-scale CMB anisotropies over a range of frequencies. A simple ray tracing diagram is shown in Fig.~\ref{fig:lens}. The design employs two roughly \SI[number-unit-product=\text{-}]{310}{\milli\meter} diameter silicon lenses with a maximum sag of about \SI{16}{\milli\meter} on the primary lens. The system has an average effective $f$-number spanning 1.5--1.7 and a telecentricity angle not exceeding \ang{0.1} over the entire field. This design supports a relatively wide diffraction-limited field of view (DLFOV) of approximately \ang{30} up to roughly \SI{350}{\giga\hertz}. This corresponds to an active focal plane area with a diameter of approximately \SI{250}{\milli\metre} (\ang{0.103}/mm plate scale) and more than 2500 physical pixels, assuming a \SI[number-unit-product=\text{-}]{6}{\milli\metre} pitch size. By employing dichroic bolometers with two polarization directions, this telescope could have over 5,000 bolometer channels.

For the purpose of studying diffraction effects as they relate to CMB polarimeters, our analysis is focused on the performance at \SI{90}{\giga\hertz}. However, the API that we have developed for the purpose of this study can be easily extended to other frequencies used to map the polarization of the CMB. 

\subsection{Unshielded telescope beams}
We generate a linearly polarized Gaussian pattern at \SI{90}{\giga\hertz} on the focal plane, with a far-field beam FWHM of \ang{16}. This corresponds to a conservative \SI{-18}{\deci\bel} edge taper on the cold stop, which should produce relatively low amplitude sidelobe response.
The pixel is polarized such that the electric field is oscillating horizontally relative to gravity, corresponding to the right panel in Figure~\ref{fig:Sommerfeld}. We do not consider the other polarization in this proceeding.

The source can be translated on the focal plane and the maximum field location (the edge of the focal plane), corresponding to roughly \ang{15} field angle, is approximately \SI{125}{\milli\metre}. 
The field from the Gaussian source is propagated through the secondary and the primary lenses using PO.
The equivalent currents on the primary can then be used as a source for PO analysis of the forebaffle.
In a real telescope, passive elements such as filters and the vacuum window could scatter light. 
We do not simulate these effects directly, instead we approximate any scattering by adding one or several unpolarized Lambertian sources on the sky side of the primary lens, at the aperture of the forebaffle. 
Testing indicated that the number of sources made very little difference for our simulations, as long as their total intensity remained constant. Therefore, all simulations use a single unpolarized Lambertian source with optical power that corresponds to \SI{0.5}{\percent} of the detector horn illuminating the secondary lens.

\subsection{Baffling the telescope}
\label{sec:baffling}
As our telescope design has a wide field of view, we need to be mindful of ground pickup, illumination from the Moon and the Sun as well as other unwanted sources of radiation. 
To suppress it, our telescope includes a conical forebaffle mounted beyond the vacuum window and a fixed reflective ground screen several meters away from the optics. 
The forebaffle is a truncated cone of length $h_{\mrm{fb}}$ and apex half-angle $\alpha_{\mrm{fb}}$. The circular aperture on the primary side is situated \SI{10}{\milli\metre} in front of the primary lens and has a radius of \SI{170}{\milli\metre}. The inner side of the forebaffle can be either perfectly reflective or absorbing. We also have the option of a \SI[number-unit-product=\text{-}]{5}{\milli\metre} thick layer of dielectric meant to mimic ECCOSORB HR-10\cite{Qubic20-8}, with a relative electric permittivity $\varepsilon_r=3.54$ and loss tangent $\delta = 0.057$.

To mitigate possible diffraction effects at the outer rim of the forebaffle\cite{Qubic20-8}, a curved flare can be added, whose radius is $n\lambda$, with $\lambda$ corresponding to the wavelength of our source.
The ground screen is, at its base, a cylinder \SI{7}{\metre} in radius. At a height $z_t$, the wall starts angling outward at an angle $\beta$ to the vertical. 
It reaches a maximum height of $z_{\mrm{apex}}$. We assume the ground level to be at the bottom of the ground screen, and the ground within the screen to be fully absorbing.
The telescope rotates in elevation (and azimuth) around the centre of the secondary lens, which is \SI{1.5}{\metre} above the ground. 

\begin{figure}
    \centering
    \includegraphics[width=.6\textwidth]{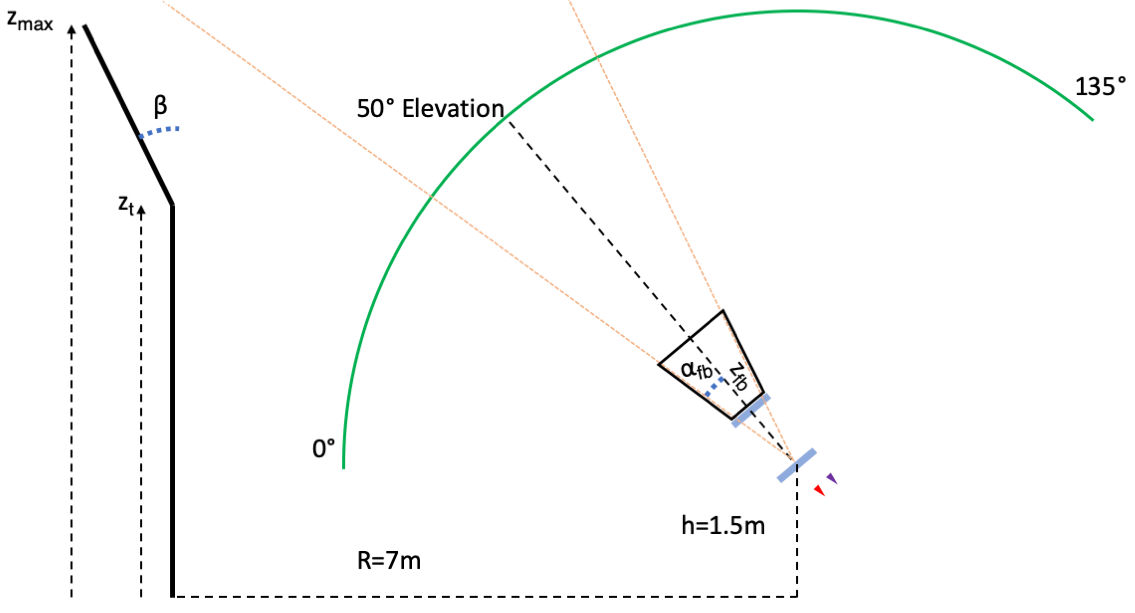}
    \caption{Schematic view of the telescope and ground screen with relevant angles distances labeled. The thick solid black line is the ground screen, the trapezoid represents the forebaffle, the blue elements are the lenses and the source pixel is the red cone. The purple cone represents a pixel at the edge of the field of view with a centroid shifted by \ang{14.5} relative to the principal ray. The elevation spherical cut is depicted here inside the ground screen for ease of view; it is in reality in the far field. The field of view is indicated by the orange dotted cone.}
    \label{fig:BaffledTelescope}
\end{figure}

\subsection{From the forebaffle to the far-field}
\label{sec:config}

The GRASP GTD simulation framework does not allow us to use the lens or the forebaffle as a direct source.
Therefore, we sample the total field generated by the main beam, the Lambertian source, and the possible scattering by the forebaffle using a near field grid placed a few centimetres beyond the aperture of the forebaffle. 
This grid is a pattern of 24 equally spaced radial segments, each comprised of 150 uniformly spaced points, extending to \SI{5}{\centi\metre} beyond the edge of the forebaffle's rim. 
We chose that density of points after testing several other configurations: under-sampling the grid would result in unphysical results. 
We allow the GTD rays to interact once with the ground screen between that source and their far-field destination, an elevation cut spanning from 0 to 135 degrees over the horizon.

We do not want the forebaffle to obstruct our telescope's field of view, but we also want it to suppress sidelobes and illumination by bright sources. 
These two goals pull the constraints on the forebaffle design in different directions.
We set $z_t=\SI{3.64}{\metre}$, $\beta=\ang{32.5}$ and $z_{\mrm{apex}}=\SI{5.64}{\metre}$. 
In this case, the ground screen's edge is at \ang{27} elevation as seen from the center of the secondary lens, and at \ang{26.5} elevation when seen from the center of the primary lens when the telescope's elevation is \ang{50}. 
It disappears from the primary's \ang{29} field of view at a telescope elevation of \ang{42}.
We model fully reflective forebaffles with $\alpha_\mrm{fb}$ either 16, 18 or \ang{20}, and three forebaffle lengths: 60, 80 or \SI{100}{\centi\metre}.
There is no curved rim on these forebaffle models.
The flare's radius of curvature is either 5, 15 or 25 times the wavelength of our main source, $\lambda \approx \SI{3.3}{\milli\metre}$. 
When a flare is implemented, we extend the forebaffle until its surface is nearly normal to the optical axis.

In the case of ground-based experiments, scan-synchronous noise caused by sidelobe coupling to warm objects such as the ground, mountains, or near-by structures can impact data analysis. In general, sidelobe response can pick up strong signal from point sources and the Galactic plane and create a residual in maps of regions thought to be relatively clean. As the amplitude of the sidelobe profile obviously impacts the detector signal, reducing the level of the sidelobe response is a key design goal of most if not all experiments observing the CMB. As an example, we note that a \SI{-50}{\deci\bel} sidelobe coupling to a 100 square degree object at \SI{300}{\kelvin} should cause $\sim\SI{0.1}{\pico\watt}$ loading on a \SI[number-unit-product=\text{-}]{90}{\giga\hertz} single-moded \SI{25}{\percent} frequency bandwidth detector coupled to a \SI[number-unit-product=\text{-}]{30}{\centi\metre} aperture. This is roughly 5\% of the loading caused by Earth's atmosphere assuming \SI[number-unit-product=\text{-}]{1}{\milli\meter} precipitable water vapor (PWV) and a \ang{45} elevation angle. Reducing the sidelobe amplitude down to -60 dB, \SI{-10}{\deci\beli} assuming a \SI{50}{\deci\beli} forward gain, reduces that signal amplitude by an order of magnitude. This is quite significant given that atmospheric loading tends to dominate loading on ground-based detectors.

\section{Results}
\label{sec:results}
\subsection{Near-field}
\label{ssec:nearfield}

\begin{figure}
    \centering
    \includegraphics[width=\textwidth]{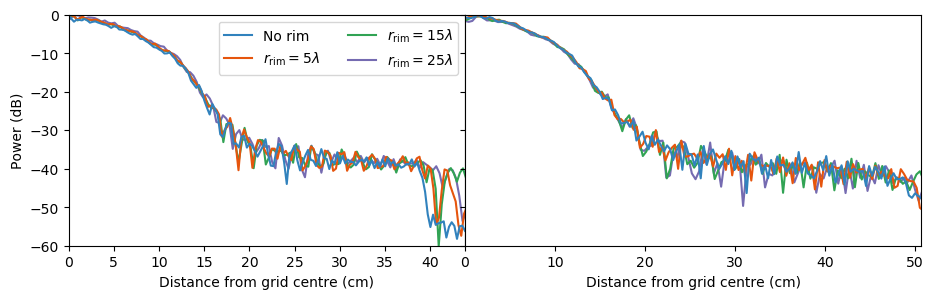}
    \caption{A cut of the near-field beam power after the edge of the forebaffle for the two geometries: on the left, $(z_\mrm{fb}=\SI{60}{\centi\metre}, \alpha_\mrm{fb}=\ang{20})$ and on the right,  $(z_\mrm{fb}=1m , \alpha_\mrm{fb}=\ang{16})$. 
    The cut's size differs from simulation to simulation, since it is defined relative to the forebaffle's aperture size and the curved rim increases it. Power is normalized so the maximum is \SI{0}{\deci\bel}.}
    \label{fig:NearFieldCurvature}
\end{figure}

\begin{figure}
    \centering
    \includegraphics[width=\textwidth]{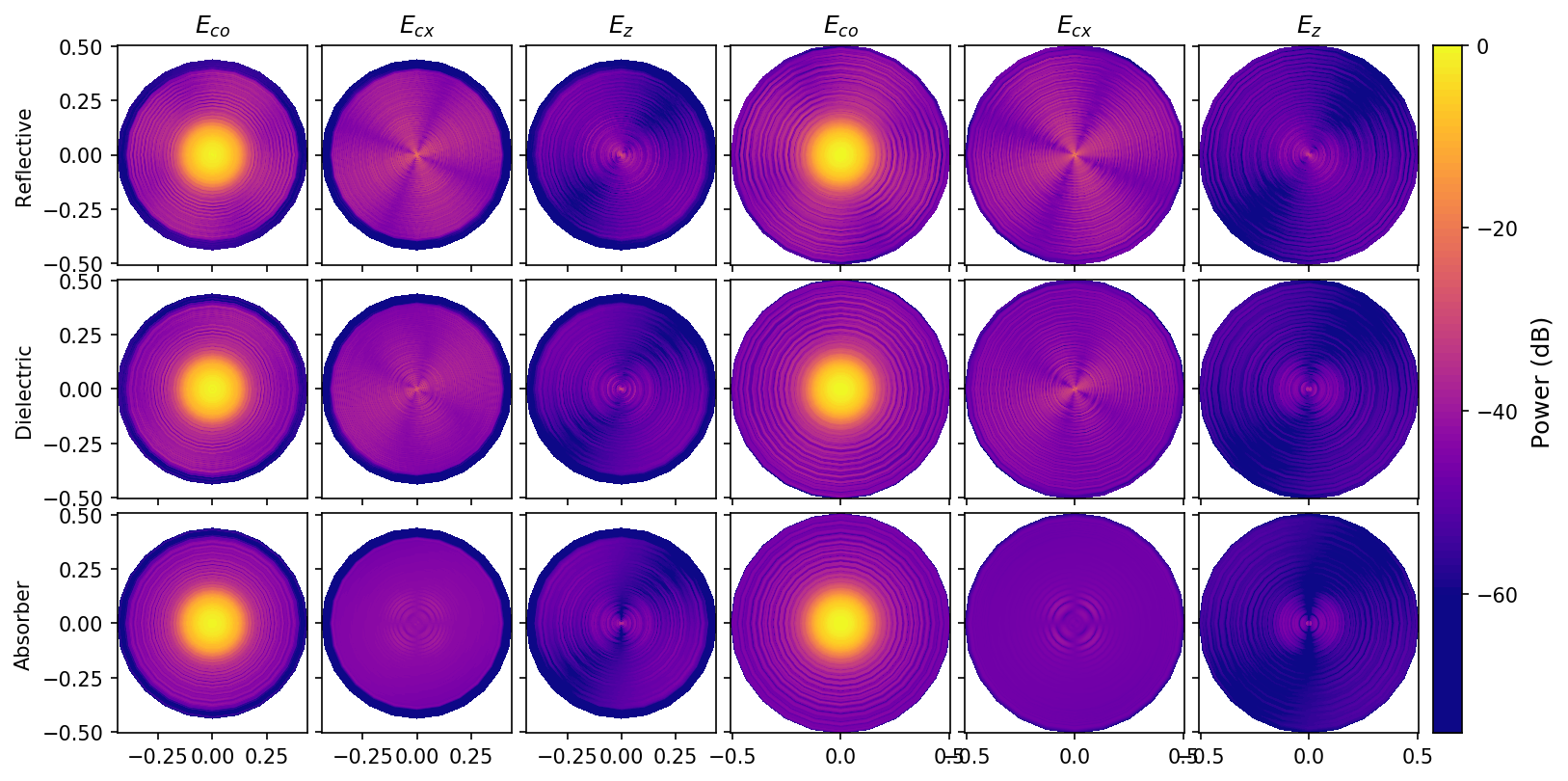}
    \caption{The components of the near-field that serve as input for a flare-less forebaffle, with different material properties in each row. The first three columns are for a forebaffle with a $z_\mrm{fb}=60$cm and $\alpha_\mrm{fb}=\ang{20}$. The last three columns are for a forebaffle with a $z_\mrm{fb}=1$m and $\alpha_\mrm{fb}=\ang{16}$. The distances on the x- and y-axes are in meters.}
    \label{fig:NearFieldMaterial}
\end{figure}

Figures~\ref{fig:NearFieldCurvature} and \ref{fig:NearFieldMaterial} show the near-field beam power patterns for a selection of forebaffle configurations, normalized to the maximum of the co-polar beam. 
All the models depicted are based on either a short forebaffle $\{z_\mrm{fb}=\SI{60}{\centi\metre}, \alpha_\mrm{fb}=\ang{20} \}$ or a longer, narrower one $\{z_\mrm{fb}=\SI{10}{\centi\metre}, \alpha_\mrm{fb}=\ang{16} \}$. 
In Figure~\ref{fig:NearFieldCurvature}, we examine the effect of an added curved rim on these two models.
We expect that curving the rim is equivalent to apodizing the aperture of the forebaffle, smoothing the transition between the beam and the surrounding space. 
We plot a cut on our grid: since our optical system has rotational symmetry around the forebaffle's axis, a single cut should represent the physics of the co-polar beam well. 
We see the apodization happen clearly for the co-polar beam of the short forebaffle, where the model with the straight rim has a steep $\sim \SI{15}{\deci\bel}$i drop around \SI{38}{\centi\metre} from the beam's centre.
That drop gets gradually attenuated in the models with curved rims: while it is still visible when $r_\mrm{rim}=5\lambda$, it becomes far less pronounced for $r_\mrm{rim}=15\lambda$ or $ 25\lambda$.
For the longer, narrower forebaffle the apodization is not immediately apparent. 
It is just about visible at the very edge of the cut, at about \SI{50}{\centi\metre} for the straight-rim configuration.
This might be due to the size of our grid:
it extends \SI{5}{\centi\metre} beyond the rim in every simulation, and
\SI{5}{\centi\metre} represents a greater fraction of the aperture size for the short forebaffle than for the long one.

In Figure~\ref{fig:NearFieldMaterial}, we go back to the straight-rim versions of the long and short forebaffles. 
We now plot all three components of the squared electric field following the Ludwig-III polarization convention\cite{Ludwig1973}.
We see the cross-polar component is weaker when the forebaffle is coated with the dielectric described in Section \ref{sec:baffling}.
When the forebaffle is modelled as purely absorbing, that component is nearly uniformly \SI{-45}{\deci\bel} for the two models. 
Conversely, a quadrupolar pattern visible in the co-polar component of the reflective forebaffles gets suppressed: for the absorbing forebaffles, the co-polar beam is very nearly rotationally symmetric.

\subsection{Far-field}
\label{sec:ff}

\begin{figure}
    \centering
    \includegraphics[width=\textwidth]{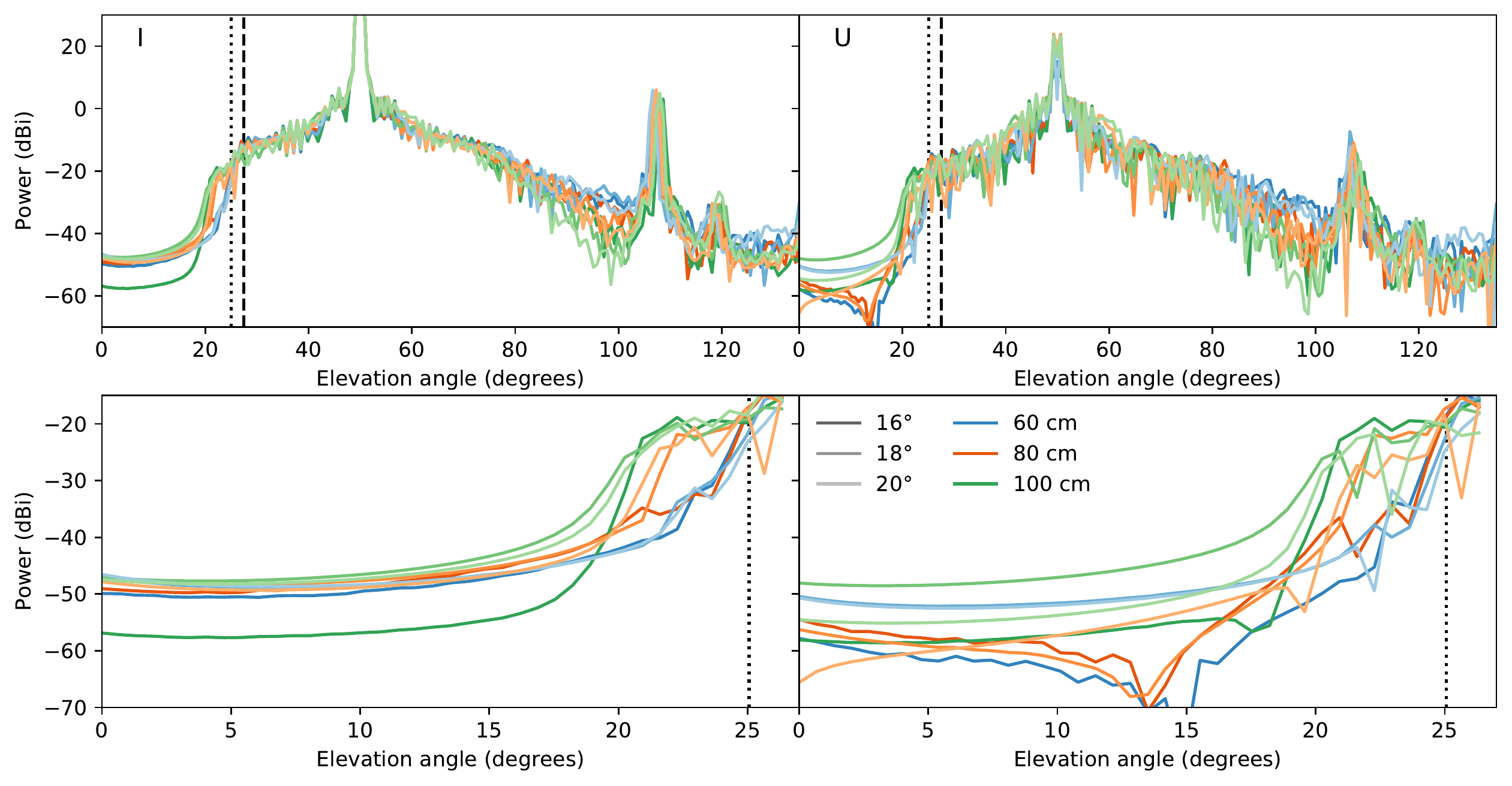}
    \caption{Elevation cut of the far-field beam predicted by GTD simulations that use tabulated fields at the forebaffle opening as input. The telescope is at $\ang{50}$ elevation. Color lightness indicates the different values of $\alpha_\mrm{fb}$. Different colors are for different forebaffle lengths, $z_\mrm{fb}$. In the top row, the dashed black line indicates the elevation of the ground screen's rim as measured from the centre of the primary lens, corresponding to \ang{26.5}. The dotted black line is the edge of the penumbra for the $\{z_\mrm{fb}=\SI{1}{m}; \alpha_\mrm{fb}=\ang{16}\}$ forebaffle. The bottom row focuses on the angular region in the ground screen's shadow and penumbra. Left column: Stokes I component. Right column: Stokes U component.}
    \label{fig:hca}
\end{figure}

Since the primary lens is finite in size, the angle from the surface of the primary to the tip of the ground screen has a finite range. We define the larger of those two angles as the penumbra and the smaller as the shadow. In some cases, the forebaffle will shield parts of the primary lens from the tip of the ground screen. In this case, the angular range of the penumbra is reduced in size. For long forebaffle configurations and/or high telescope elevation angles, no part of the primary lens will see the tip of the ground screen. In that case, the shadow is cast by the forebaffle.

We set the telescope's elevation to \ang{50}, and use the near-field patterns we described in Section \ref{ssec:nearfield} as input for our GTD simulations.
Sampling the elevation cut defined in Section~\ref{sec:config} will help us evaluate the sidelobe amplitude for different optical configurations.
We are aware that the GTD method implemented by GRASP creates caustics at the location of the main beam and we ignore this part of the simulation result. 
Instead, we use the results of a physical optics model without the ground screen in a $\pm\ang{3}$ region around the main beam.
In all results, we normalize power by the forward gain of a Gaussian beam with a FWHM set by the theoretical diffraction limit of the \SI[number-unit-product=\text{-}]{30}{\centi\metre} aperture stop; roughly \SI{50}{\deci\beli} at \SI{90}{\giga\hertz}.

\begin{figure}
    \centering
    \includegraphics[width=\textwidth]{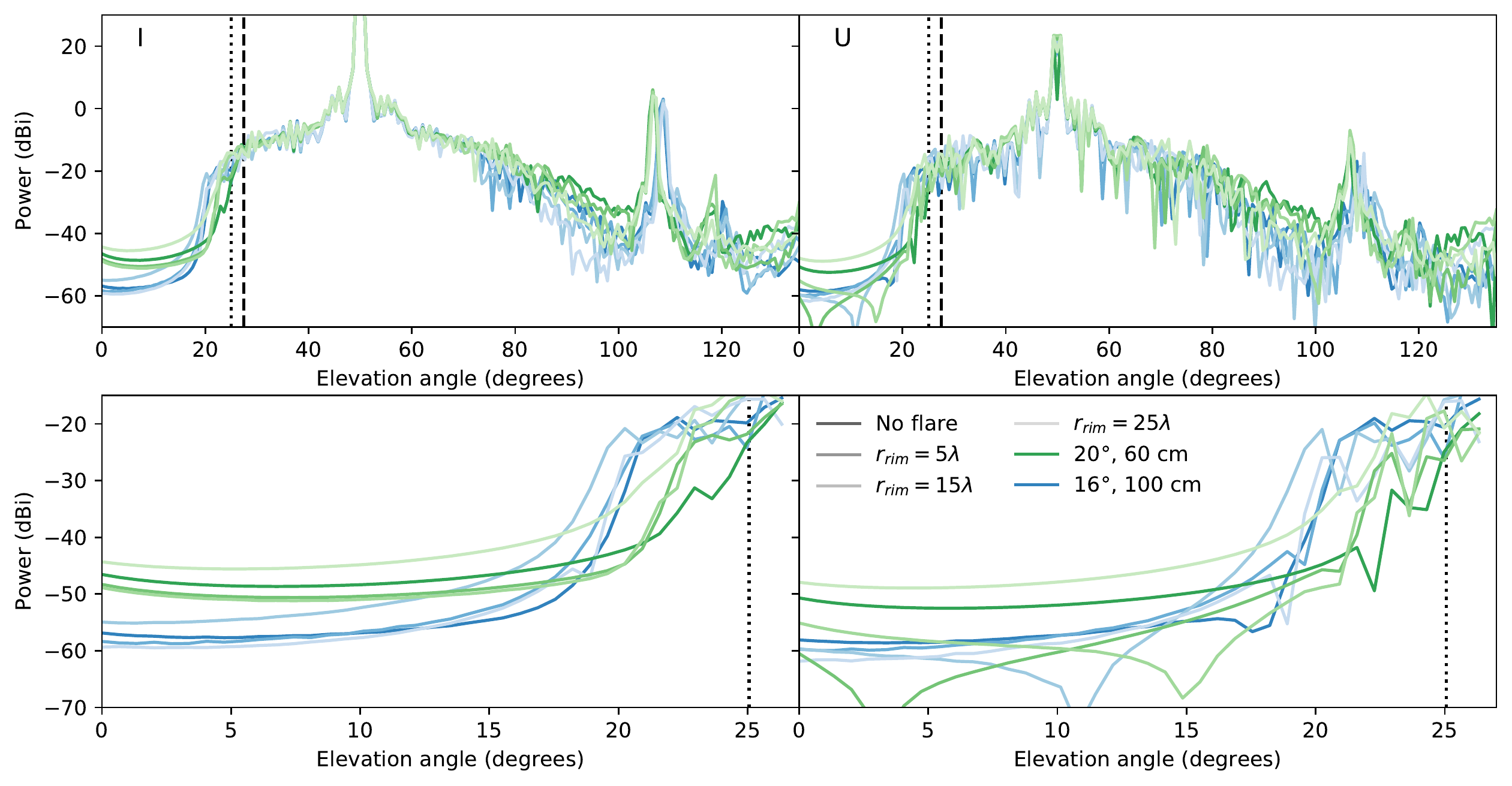}
    \caption{Elevation cuts of the far-field beam for the $\{h_\mrm{fb}=\SI{100}{\centi\metre};\alpha_{\mrm{fb}}=\ang{16}\}$ and $\{h_\mrm{fb}=\SI{60}{\centi\metre};\alpha_{\mrm{fb}}=\ang{20}\}$ forebaffles with different flared rim configurations represented by different shades of the same color. The telescope is set at \ang{50} elevation. Left column: Stokes I component. Right column: Stokes U component.}
    \label{fig:fb_curv}
\end{figure}

We start by examining the impact of the forebaffle's length and opening angle.
In Figure~\ref{fig:hca}, we plotted the Stokes I and U components of the beam for a \ang{50} elevation, with the bottom two panels focusing on the ground screen's shadow.
Surprisingly, we find that lengthening the forebaffle does not reduce the Stokes I component in the ground screen's shadow and penumbra. 
However, increasing the opening angle typically results in a larger sidelobe response. 
Conversely, for high elevation angles the long forebaffle and wide opening angles actually helps reduce the Stokes I and U sidelobe amplitude (see 70-\ang{100} region in Figure~\ref{fig:hca}). This behavior is more consistent with our  expectations.

We find that increasing the length of the forebaffle results in a translation of the drop-off position in the ground screen shadow. This could be explained by the fact that the forebaffle solid angle subtended from the ground screen is larger for a long forebaffle. The Stokes U beam profile is larger for the longer forebaffle configuration. This is caused by reflections on the forebaffle which are polarization dependent. Finally, we note a secondary \SI{5}{\deci\beli} peak in the Stokes I beam profiles at a roughly 106-\ang{108} elevation, which is due to reflection of the beam off the angled section of the ground screen. The Stokes U beam also shows a peak at this location, but the amplitude is roughly \SI{-10}{\deci\beli}.

Selecting the configuration $\{h_\mrm{fb}=\SI{100}{\centi\metre};\alpha_{\mrm{fb}}=\ang{16}\}$, hereafter the long forebaffle, and $\{h_\mrm{fb}=\SI{60}{\centi\metre};\alpha_{\mrm{fb}}=\ang{20}\}$, the short forebaffle, for further developments, we look at the impact of a flared rim.
Figure~\ref{fig:fb_curv} shows that the flare adds some Stokes I beam power at elevations between \ang{20} and \ang{25} for the short forebaffle. 
The difference is less significant for the long forebaffle. 
In the Stokes U beam component, we observe no clear benefit in adding a flare to the model, with a possible exception of the $25\lambda$ flare for the short forebaffle configuration; in which the beam response is larger.
As in the previous case. the largest difference between the flare and the no flare case can be seen at 70-\ang{100} elevation angles. In this region, we note a significant reduction in sidelobe power for both the Stokes I and U beam profiles when a flare is added. 

\begin{figure}
    \centering
    \includegraphics[width=\textwidth]{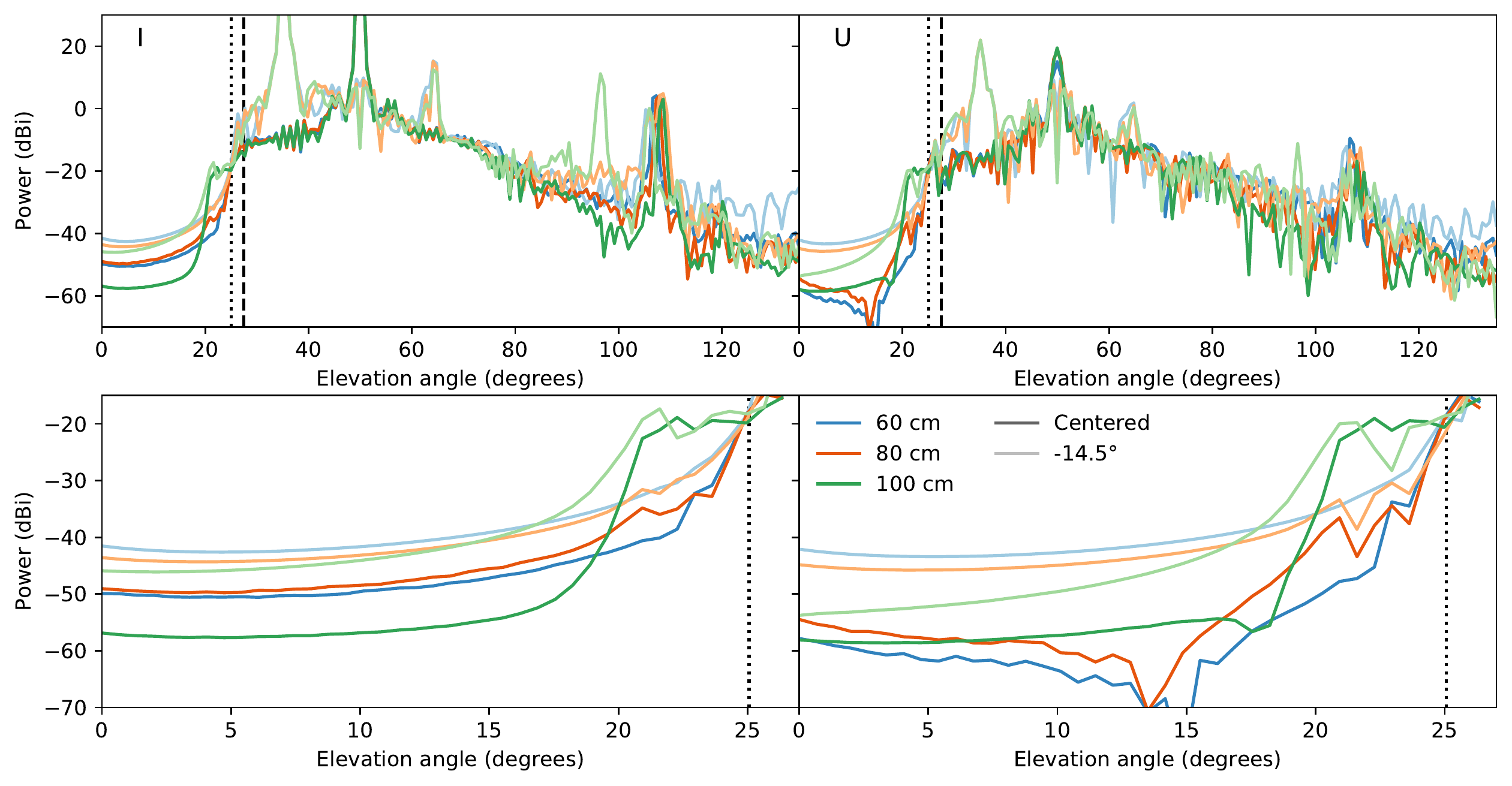}
    \caption{Elevation cut of the far-field for a forebaffle with $\alpha_\mrm{fb}=\ang{16}$. The telescope is at $\ang{50}$ elevation. Color shade indicates the different positions of the source on the focal plane. Different colors are for different $z_\mrm{fb}$.}
    \label{fig:offset_50}
\end{figure}

What happens for off-axis pixels?
We now displace our focal plane source by \SI{12}{\centi\metre} relative to the center pixel, in the same plane as the elevation cut (see purple source in Figure~\ref{fig:BaffledTelescope}).
Given the plate scale, this corresponds to a beam centroid that is offset by \ang{14.5} 
relative to the boresight. 
In Figure~\ref{fig:offset_50}, we compare results for the centered and off-center sources. 
We show the Stokes I and U beam profiles for the forebaffle lengths previously considered, with $\alpha_\mrm{fb}=\ang{16}$.
As expected, we observe that changing the position of the source affects the diffraction profile of the Stokes I beam in the ground screen's shadow, with a roughly $\sim\SI{10}{\deci\beli}$ higher power for the edge pixel. 
If some elevated terrain, such as a structure or a mountain, were to be situated in that part of the far-field, then the detector response while scanning would be significantly different for the two pixels. 
Outside the shadow, we see that the off-center pixel has a mirror image on the other side of the boresight of our telescope caused by reflections in the forebaffle.
Furthermore, for the \SI[number-unit-product=\text{-}]{1}{\metre} long forebaffle the secondary peak at \ang{105} elevation has an echo at \ang{95}.

\begin{figure}
    \centering
    \includegraphics[width=\textwidth]{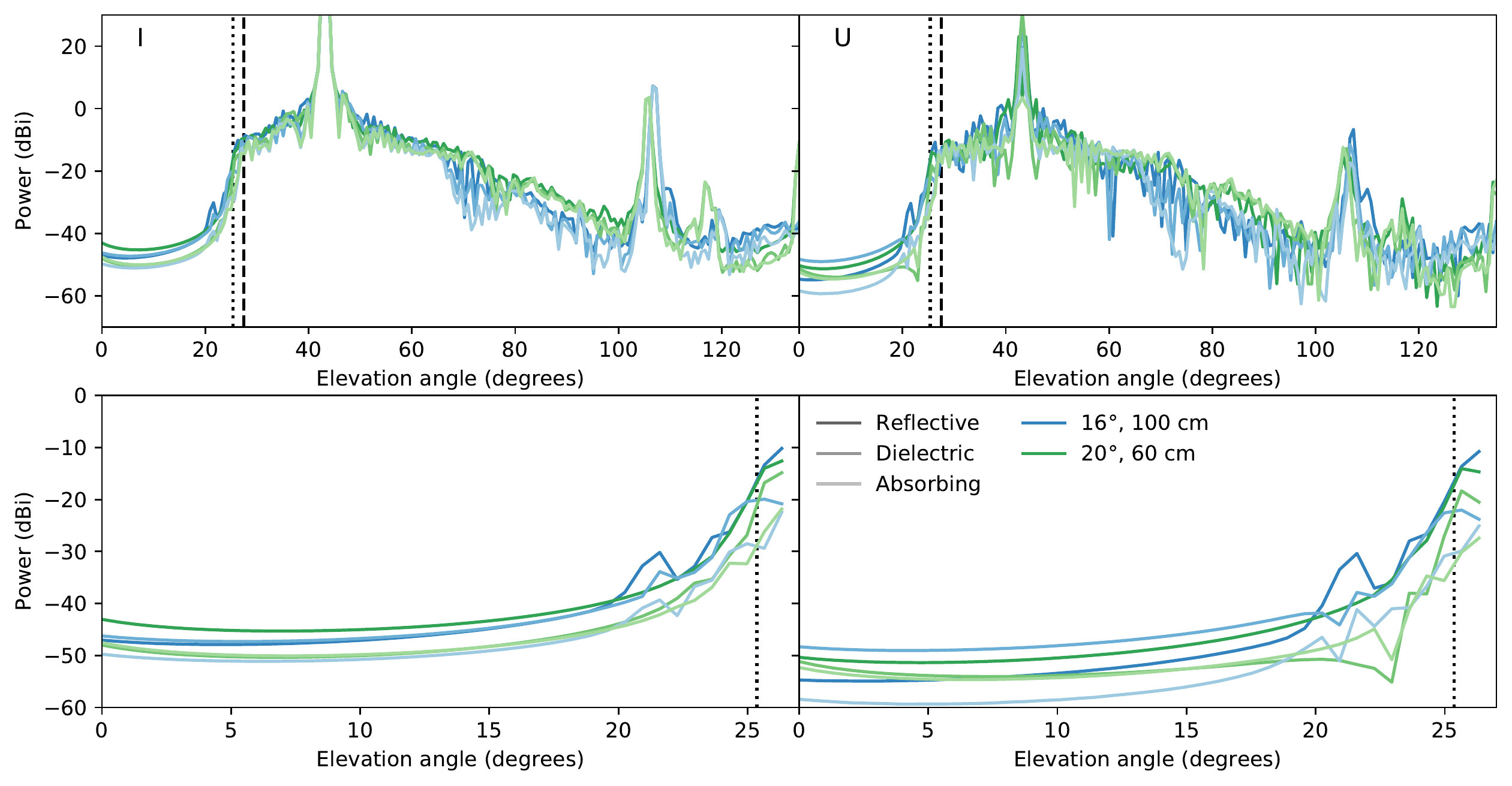}
    \caption{Elevation cut of the far-field for the long and short forebaffle models, respectively in blue and green. Different shades of the color denote different material properties. The telescope is at \ang{43} elevation.}
    \label{fig:mat}
\end{figure}

In Figure~\ref{fig:mat}, the telescope's elevation is set to \ang{43}.
We are interested in the effect of forebaffle material properties on the sidelobe response, as previously mentioned in Section~\ref{sec:baffling}. 
As expected, the power outside the main lobe is generally reduced for the dielectric and fully absorbing models relative to the reflective forebaffle.
This is true both in the ground screen's shadow and at high elevations.
The only exception is for the Stokes U beam of the dielectric long forebaffle, which is greater under \ang{20} than for the reflective model. 
The height of the secondary peak at \ang{106}-\ang{110} doesn't change significantly, confirming that it is mainly due to direct reflection of the main beam on the ground screen.

\begin{figure}
    \centering
    \includegraphics[width=\textwidth]{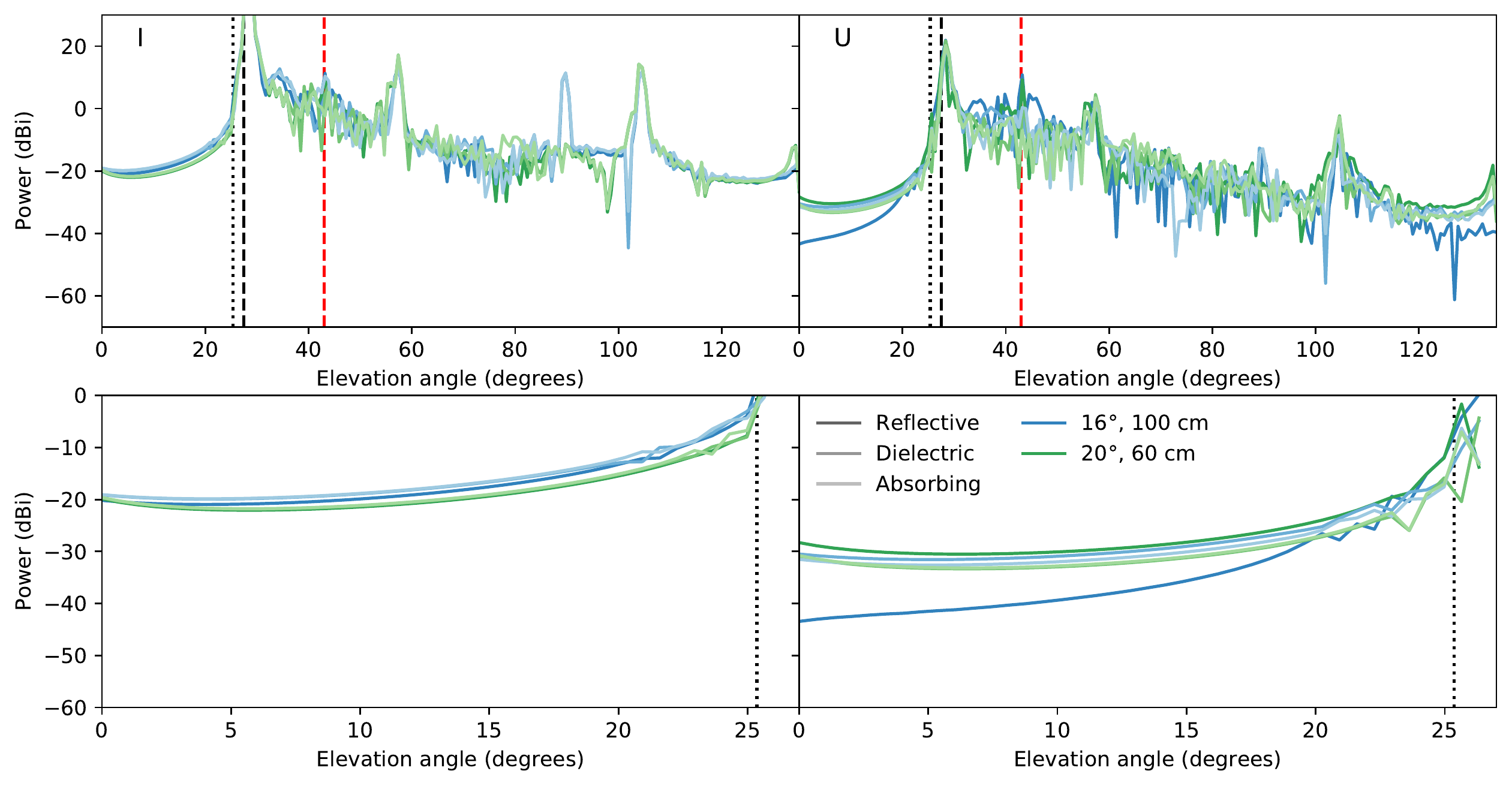}
    \caption{Elevation cut of the far-field for the long and short forebaffle models with an off-center pixel, respectively in blue and green. Different shades of the color denote different material properties. The telescope is at \ang{43} elevation, indicated by the red dashed line. The off-center detector's main lobe nearly clips the ground screen's edge. }
    \label{fig:mat_offset}
\end{figure}

When the telescope is at \ang{43} elevation, an edge pixel on the telescope will come close clipping the ground screen. 
We now repeat our simulations with an off-center pixel. 
The results are visible in Figure~\ref{fig:mat_offset}. 
Like in Figure~\ref{fig:offset_50}, we see that the diffracted power in the ground screen's shadow is greater than for the center pixel, but the effect is stronger this time with a $\sim\SI{25}{\deci\beli}$ difference in the sidelobe response for the center and edge pixels. 
In this scenario, the absorbing configuration no longer reduces power at high elevations and we also see significant differences in behavior between the short and long forebaffles. 
In the ground screen's shadow, the Stokes U beam for the reflective long forebaffle is much weaker than for the two other configurations. 
There is an additional feature at \ang{90} that is only visible for the long forebaffle. The cause of this feature is currently unknown.

\section{CONCLUSION}
\label{sec:conclusion}
Optical modeling of CMB telescopes can be done through several complementary methods.
In this paper, we introduced a design for a simple ground-based telescope that could map the cosmic microwave background on degree angular scales.
We have then used physical optics to simulate the near-field of primary optics and forebaffle of this telescope at \SI{90}{\giga\hertz}. 
Using GTD, we then examined the effect of a ground screen for varying forebaffle properties. 
We found that varying the length and opening angle of the forebaffle had opposite effects at different elevations, with longer and wider forebaffles increasing the sidelobe response in the shadow and decreasing it at high elevations.
For the design considered, a flared rim on the forebaffle was not found to reduce the power in the ground's screen shadow. 
However, we found that it reduces the power at high elevations. 
As expected, off-center pixels have a different beam profile in the shadow compared to centered ones. 
This implies an elevation dependent variation in scan-synchronous noise for ground-based telescopes.
The beam profile of off-center pixel is also found to have secondary peaks due to reflection off the forebaffle.
Finally, we see that making the forebaffle absorbing reduces sidelobe power for a centered pixel, but will have varying effects for an off-center pixel depending on the elevation. 

Although the GTD simulation framework provided by GRASP is unable to capture diffraction on curved edges, we find that this software is still able to produce useful models of the far-sidelobe response for a generic experiment designed to map the degree-scale polarization of the CMB.
These results are all reported for a polarization perpendicular to the plane of incidence as shown in Figure~\ref{fig:Sommerfeld}.
In future work, we will extend and adopt this modeling framework to real designs of upcoming CMB experiments.

\acknowledgments 
 
We thank Lyman Page for useful discussions at the beginning of this project. We would also like to express our gratitude to Per Heighwood Nielsen, Senior Research Engineer at TICRA, for his assistance in understanding the capabilities and limitations of the software.

\bibliography{main} 
\bibliographystyle{spiebib} 

\end{document}